\documentclass{jltp}
\usepackage{latexsym}
\usepackage{amssymb}
\usepackage{amsmath}
\usepackage{graphicx}
\newcommand{\hef}{$^4\mathrm{He}$}

\begin{document}

\title{Trapping electrons in electrostatic traps over the surface of \raisebox{0.8
    ex}{\bf{\footnotesize 4}}He.}
\address{\hskip -1.5ex CEA-Saclay/DSM/DRECAM/ SPEC, 91191 Gif sur Yvette, Cedex, France\\
  \hskip -1.7ex$^*$Present address: Chemistry Dept.,\! North Carolina State
  Univ.,\! Raleigh,\! NC 27695
}
\author{\hskip -1.5ex E.\! Rousseau,\! Y.\! Mukharsky, D.\!
  Ponarine$^*$\hskip -0.5ex, O.\! Avenel, E.\! Varoquaux}
\date{\today}

\runninghead{E. Rousseau et al.}{Trapping electrons in electrostatic traps over
  the surface of helium}

\maketitle

\begin{abstract}
  We have observed trapping of electrons in an electrostatic trap formed over
  the surface of liquid \hef.  These electrons are detected by a Single Electron
  Transistor located at the center of the trap.  We can trap any desired
  number of electrons between 1 and $\sim 30$.  By repeatedly ($\sim
  10^3-10^4$ times) putting a single electron into the trap and lowering the
  electrostatic barrier of the trap, we can measure the effective temperature of
  the electron and the time of its thermalisation after heating up by
  incoherent radiation.

\end{abstract}

In 1999, Platzman and Dykman\cite{Platzman:1967} proposed that single
electrons electrostatically trapped on the surface of a liquid helium film
could be used as qubits and hence form the basis of a quantum computer. This
proposal quickly aroused experimentalists'\,interest.\cite{Lea:2000} Here we
present experimental results on the trapping of a single (or other desired
number) electron and measurements of the upper limit of its relaxation time in
view of defining its usefulness as potential qubit.

\section{The experiment}

\begin{figure}[t]
  \centerline{\mbox{
    \includegraphics[height=4.2cm]{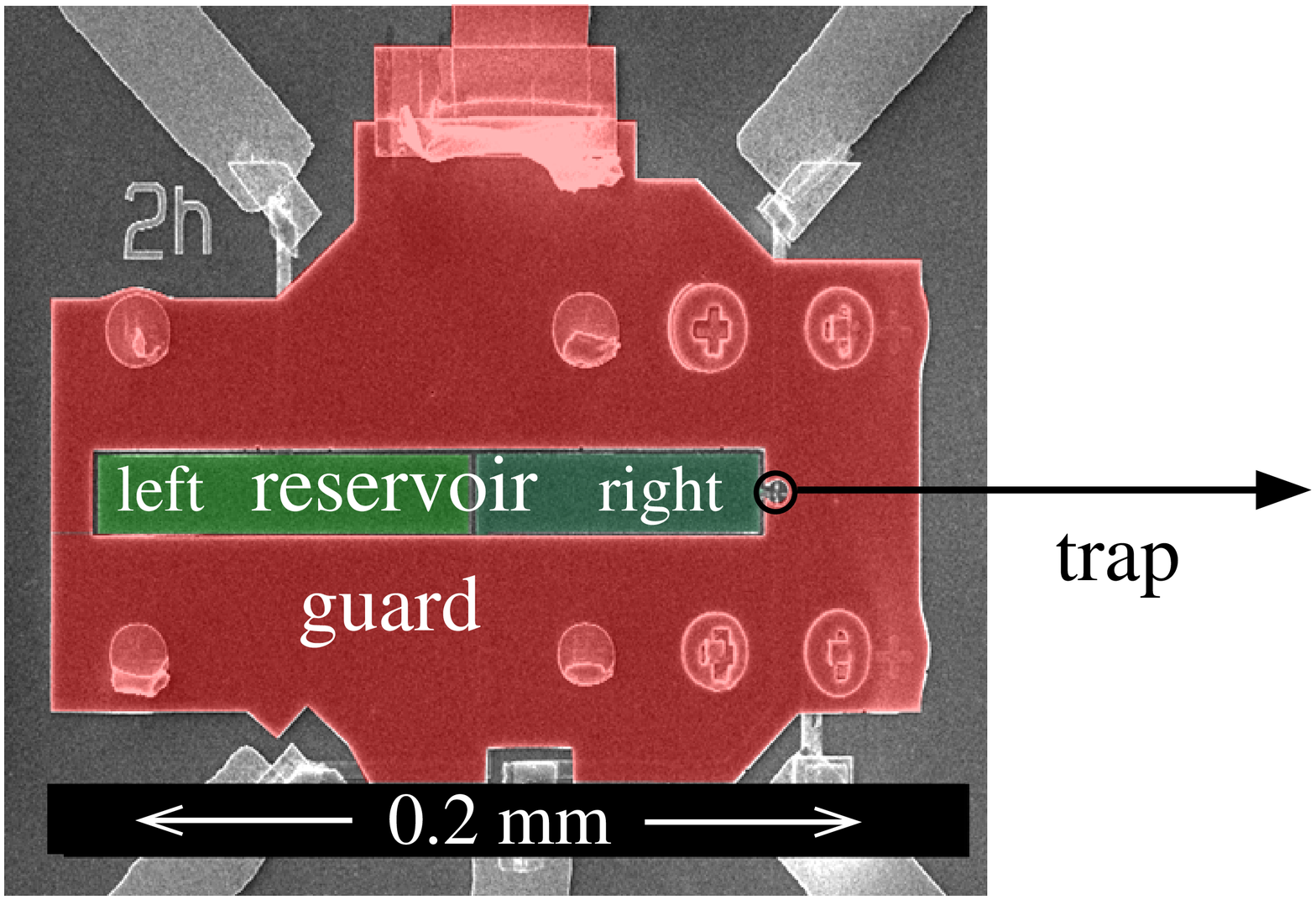}
    \hskip -0.2 cm
    \includegraphics[height=4.25cm]{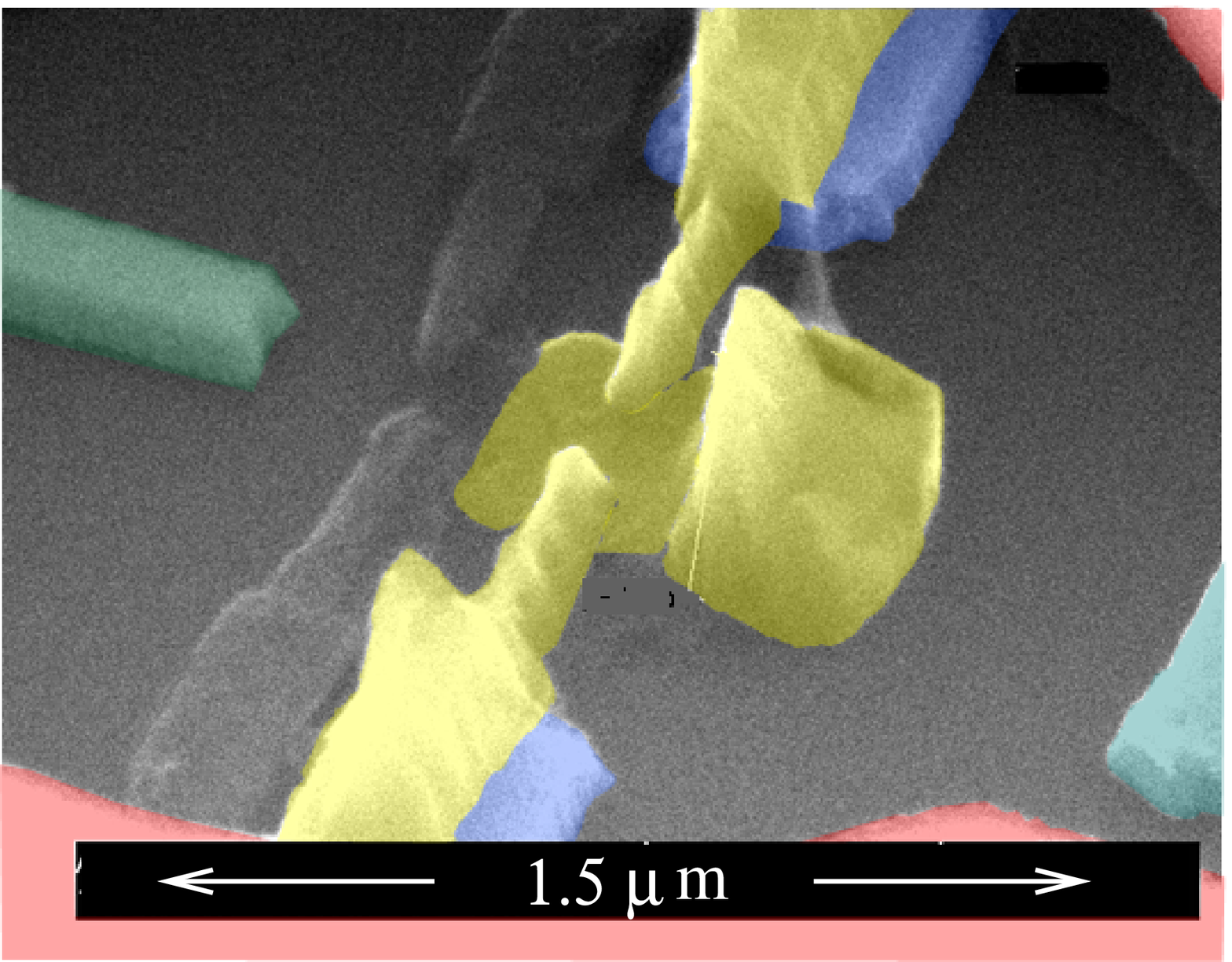}
  }
}
\caption{
\label{sample}
Scanning Electron Microscope photograph of the microfabricated 
part of the sample (left) and the electron trap (right). (color online)
}
\end{figure}

The tested device consists in a system of electrodes designed to hold an
electron in a well-defined position over the liquid helium surface. A single
electron transistor (SET) located below the electron is used to monitor its quantum
state.  The whole structure is represented in Fig.\,\ref{sample}. It comprises an
electron reservoir and the electron trap, which is supplied in electrons from
the reservoir, and a guard electrode made out of a thick ($\sim $0.5~$\mu $m)
layer of Nb. The structure is covered by a saturated helium film $\sim
$200-400~{\AA} in thickness. Two electrodes, made out of thin Niobium and lying at
the bottom of the reservoir, fix its electrostatic potential with respect to
the guard.

A gorge through the guard electrode connects the reservoir to the trap. When
this electrode is negatively biased relative to the SET, a
potential barrier forms at this gorge, isolating the trap from the reservoir.

The voltage across the SET is modulated by the charge on its island, $q_I
=\sum_i {C_i U_i } +\sum_j {q_{0,j} } $. The first sum is over all the
conductors in the systems having capacitance $C_{i}$ and potential $U_{i}$
with respect to the island.  The second sum is over the image charges induced
on the island by stray charges in the rest of the system, such as charges
or dipoles in the substrate and the electrons over helium. The voltage
$V_{s}(q_{I})$ is \textit{a priori} unknown. It depends on the bias current
and changes from run to run. However, it must be periodic with period $e$, the
charge of the electron.

To reduce the noise due to fluctuations of the \textit{dc} voltage across the
SET, we apply a low-frequency (80-150~Hz) modulation of the order of 100 $\mu
$V to the guard electrode, which is strongly coupled to the SET island.  The
voltage variation across the SET produced by this modulation is detected by a
lock-in amplifier. The amplitude of this signal $V(q_{I})$ is proportional to
the derivative of $V_{s}(q_{I})$ function and, thus, is periodic itself.  We
determine $V(q_{I})$ experimentally by sweeping the potential, $U_{e}$, of one
of the electrodes, usually the one which will be swept in subsequent
measurements. We then choose a part of the response to the sweep where
background charges did not change, so that several periods of the function
$V(q_{I})=V(U_{e})$ can be superimposed by transforming $U_{e}$ modulo $P$,
where the period $P$ is adjusted self-consistently. After the modulo
transformation the chunk of data is averaged and interpolated by a smoothing
spline function $\mbox{spl}(U_{e})$. The rest of the data is fitted piecewise
with the function $A\cdot\mbox{spl}(U_e+q P)$, where the amplitude $A$, and
the phase $q$ are the fitting parameters.
Parameter $q$ is the charge, expressed as fraction of $e$, 
induced on the SET island by stray charges in the system. 


An electron seeding and monitoring experiment proceeds as follows.

\noindent 
1.~We seed the electrons on the helium surface by igniting a corona discharge 
in a small chamber separated from the rest of the cell by a transmission 
electron microscope sample grid.
The cell is heated to $\sim $1.1~K before a discharge. 
The presence of the electron normally is easily detected by applying a voltage
$U_{exc}\sim $100 $\mu $V at 100~kHz to the right reservoir
electrode and measuring with a lock-in amplifier the voltage induced on the
left electrode.  When electrons appear on the surface, the signal changes by
10 to 200 nV rms.

\noindent
2.~After electrons are seeded, we let the system cool down from 1~K and
proceed to trap the desired number of them in the trap. Typically, at this
point, the guard electrode is biased to a negative potential, between $-0.1$
and $-0.5$~V, and the SET is biased to a positive potential between 0 and
$0.5$ V. First, we lower the voltage applied to the reservoir electrodes
to charge the trap. Care should be taken not to lower the voltage too much,
since this can cause irreversible loss of all the electrons. Normally, we find the
suitable potentials by trial and error. In typical cases, at least one of the
reservoir electrodes stays more positive than the guard, although the
electrons can be kept even with both electrodes negative with respect to the
guard.



\begin{figure}[tbp]
\centerline{
\includegraphics[height=5 cm]{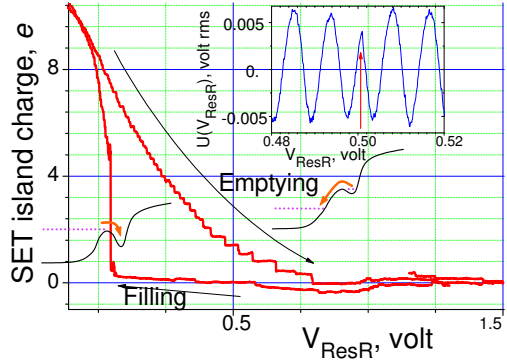}
}
\caption{
\label{Charging}
Reduced charge on the SET island in terms of the potential of the right
reservoir electrode.  SET potential is +0.6 V, Guard electrode potential
0. The bottom insert schematically shows calculated profile of the potential
along the centerline of the sample, the dashed line schematically represents
local energy of the electrons.  The top insert shows small portion of a raw
SET signal containing one of the jumps (indicated by the arrow). (color
online) }
\end{figure}

\noindent
3.~After the trap is charged, we start ramping up the potential of the right
reservoir electrode. Typical results are shown in Fig.~\ref{Charging}, where
we show the dependence of the phase of the SET signal oscillations on the
voltage applied to that electrode. When we sweep the voltage from large
positive values down, the charge induced on the SET changes little, until we
reach the charging voltage. Small changes seen in Fig.~\ref{Charging} are
probably due to motion of charges and dipoles in the substrate and occur even
before the electrons are seeded for the first time after cooldown from room
temperature.  At the charging voltage ($\sim $0.26~V in the figure) the SET
detects a sudden change in the charge. Since a SET measures charge modulo $e$,
we can not determine the real change and adjust it by integer number of $e$ to
``close the loop''.

The scheme of distribution of potentials at the moment of charging is shown in
the insert of Fig.~\ref{Charging}. The solid line represents the potential
created by the electrodes along the center of symmetry of the sample in the
plane of the electrons. The dashed line indicates the energy of the
electrons, which, due to Coulomb repulsion, is higher than this potential by
$ned/\varepsilon \varepsilon _{0}$, where $n$ is the electron number surface
density and $d$ is their distance from the electrode.

When ramping down the potential, the energy of the electrons over
the reservoir drops, but electrons in the potential well stay trapped.  As the
voltage on the right reservoir electrode is raised further, the barrier height
decreases and electrons start leaking out one by one.  Each electron
leaving the trap monitored on the SET signal as a sudden change in the phase of
the oscillations, shown in the top insert. Finally, at $\sim $0.63~V the last
electron leaves the trap. Between $\sim $0.57 and $\sim $0.63~V
only one electron is left in the trap.

The amplitude of the jump depends on the number of the electrons residing in
the trap and goes to $\sim 0.4 e$ for a few electrons. This value is to be
compared with $\sim 0.5 e$ obtained from the numerical simulation. The length
of the plateaus between the jumps depend on the Coulomb repulsion between the
electrons.

These results are rather similar to the results of the Royal Holloway
group\cite{Glasson:2005}.  However, in the present experiments, the steps are
more stable and better defined, being both ``higher'' and ``longer'', because
of the enhanced coupling to the pyramidal SET, compared with the flat-island
SET and the smaller trap used in London.

\section{Electron escape}

\begin{figure}[t]
  \begin{center}
     \includegraphics[height=5 cm]{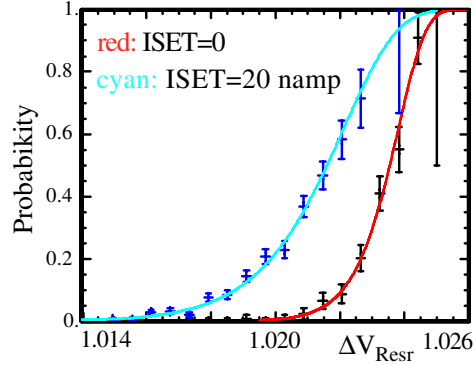}
     \caption{
     \label{Escape}
     The probability of the escape of a single electron from a trap, 
     in terms of the amplitude of the pulse applied to the right reservoir. 
     Error bars and fits are discussed in the text. $V_{set} = +0.6$~V,
     $V_{gate} = +0.0$~V, $V_{guard} = -0.1$~V.  
     The calculated potential barrier, used in the fit is
     $(0.16\cdot\mbox{volt})(V_{ResR}/\mbox{volt}-1.042)^{1.6}$. (color online) 
     }
  \end{center}
\end{figure}

Having developed the technique of trapping the electrons, we proceeded to the  
measurement of their escape from the trap.
Having prepared the trap with one electron, we momentarily open it partially
by applying short ($\sim 0.1 \mu$s) pulses to the right reservoir
electrode.  We then examine the SET signal to determine the presence of the
electron. If it has escaped, we re-populate the trap and repeat the
measurement. After accumulating $\sim 1000$ attempts, we plot the resulting
electron escape probability \textit{vs.} the amplitude of the pulse opening
the trap, as shown in Fig.~\ref{Escape}. The error bars on
the figure come from the statistical analysis of the data.

The dependence of the escape probability from a potential well on some
parameter can be calculated as in Ref.[\onlinecite{Varoquaux:2003}]:
$p=1-\exp(-\mathit\Gamma\tau e^{-E(U)/kT})$, where $\mathit\Gamma $ is the
attempt frequency of the escape, $\tau $ is the pulse length and $E(U)$ is the
height of the energy barrier, which depends on parameter $U$.  This
formula applies also to the case of quantum tunneling through a barrier if it
can be approximated by cubic+parabolic potential. In this case one should use
the effective temperature $k_BT_q=\hbar\omega_t/2\pi$, where $\omega_{t}$ is the
frequency of the oscillations of the particle in the well.

We have calculated the barrier height in terms of the voltage on the right
reservoir numerically, using the known sample geometry. To account for
possible contact potentials between electrodes, we have assumed that at the
lowest temperature the escape of the electron is due to quantum tunneling.
Then the calculated probability of escape has been fitted to the data with the
SET potential as the fitting parameter. Here, we did not make assumption about
the shape of the potential; calculated profiles of the potential barrier have
been used. Using this adjusted value of the SET potential, the $E(V_{ResR})$
function is computed, which is then used in the subsequent fits.  The fits are
quite insensitive to the values of the attempt frequency and the pulse length,
so we have used the fixed numbers of 40~GHz and $0.1\ \mu$s,
correspondingly.

\noindent
\section{The electron temperature} 

The escape curve changes as a function of the current through the SET. To make
this measurement, we apply the desired current to the SET, apply the measuring
pulse to the right reservoir and then change the SET current to a value
suitable for measurements ($\sim $3~nA). When the current through the SET is
increased, the curve shifts to the left and becomes less steep. Both effects can be
explained by an increase of the electron temperature. Such an increase is not
surprising since the current through SET generates an electric field in its
vicinity as electrons tunnel into and out of the island, changing the
potential of the island by $\sim e/(2 C_\Sigma)$, where $C_{\Sigma }$ is the
total capacitance of the island.  Knowing the barrier, we can calculate the
temperature of the electron from the fitting parameters. We find that the
temperature of the electron at zero current through the SET is 270~mK,
rising to 0.4~K at $I_{SET}=20$~nA while the cell temperature is $\sim\,$13
mK. Finally, when raising the cell temperature above 270 mK, we observe that
the temperature of the electron increases more slowly than that of the cell.  

Also, we tried to estimate the thermalisation time of the electron. After
heating the electron by $I_{SET}$ we turn off the current and, after a delay,
apply the measuring pulse. We determine the electron temperature by repeating
the measurement a number of times and using the procedure outlined above.
We find that the temperature always returns to a stationary value even for the
shortest time delay that we could apply of $\sim 0.5 \mu$s.

Both the anomalous behavior of the electron temperature and its short
thermalisation time indicate a rather strong coupling with parts of the
environment. We suspect that these features are due to the strong coupling
with the pyramidal SET and by the large pressing field acting on the
electrons.


\begin{thebibliography}{2}
\bibitem{Platzman:1967} P.M.Platzman and M.I.Dykman, {\it Science} \textbf{284},
  1967 (1999).
\bibitem{Lea:2000} M.J. Lea, P.G. Frayne, and Yu. Mukharsky,
  {\it Fortschr. Phys.} {\bf 48}, 1109 (2000).
\bibitem{Glasson:2005} P. Glasson, G. Papageorgiou, K. Harrabi, et al., \textit{J. Physics and Chemistry of Solids }\textbf{66}, 1539 (2005).
\bibitem{Varoquaux:2003}E. Varoquaux and O. Avenel, \textit{Phys .Rev. B} \textbf{68}, 054515 (2003).
\end{thebibliography}
\end{document}